\documentclass[letterpaper, 10 pt, conference]{ieeeconf}  

\IEEEoverridecommandlockouts                              
\usepackage{graphicx}     
\usepackage{url} 
\usepackage{algorithm}
\usepackage{algorithmic}
\usepackage{amsmath,amssymb,amsfonts}
\usepackage[font=small,labelfont=bf]{caption}
\usepackage{color}
\usepackage{cite}

\usepackage{pgf}
\usepackage{tikz}
\usetikzlibrary{arrows,automata,positioning}

\usepackage{amsthm}
\newtheoremstyle{exampstyle}
  {0.1cm} 
  {0.1cm} 
  {\it} 
  {} 
  {\bfseries} 
  {.} 
  {.5em} 
  {} 

\theoremstyle{exampstyle} 
\theoremstyle{exampstyle} 
\theoremstyle{exampstyle} 
\theoremstyle{exampstyle} 
\theoremstyle{exampstyle} 
\theoremstyle{exampstyle} \newtheorem{definition}{Definition}
\theoremstyle{exampstyle} 
\theoremstyle{exampstyle} \newtheorem{problem}{Problem}
\theoremstyle{exampstyle}

\newcommand{\T}{\mathcal{T}} 
\newcommand{\A}{\mathcal{A}} 
\newcommand{\B}{\mathcal{B}} 

\renewcommand{\P}{\mathcal{P}} 
\newcommand{\init}{\mathit{init}}

\newcommand{\sync}{\mathit{sync}}

\newcommand{\stay}{\mathit{stay}}
\newcommand{\Sync}{\mathit{Sync}}
\newcommand{\TS}{{\T=(S,s_{\init},{A} ,T, \AP, L)}}
\newcommand{\TSi}{{\T_i=(S_i,s_{\init,i},{A_i} ,T_i, \AP_i, L_i)}}
\newcommand{\Lab}{\mathcal{L}}

\newcommand{\irrelevant}{\Epsilon_i}
\newcommand{\Dep}{\mathit{Dep}}
\newcommand{\trans}{\mathfrak{t}}

\newcommand{\LTLX}{LTL$_{\setminus \Next}$ }

\newcommand{\N}{\mathcal{N}}

\newcommand{\model}{\mathcal{M}}
\newcommand{\I}{\mathcal{I}}

\newcommand{\Ser}{\Sigma} 
\newcommand{\Sers}{\mathbf{\Sigma}}
\newcommand{\AP}{\Pi} 

\newcommand{\Set}{\mathsf{S}} 

\newcommand{\Epsilon}{\mathcal{E}}

\renewcommand{\i}{\iota}

\newcommand{\Nat}{\mathbb{N}} 
\newcommand{\Real}{\mathbb{R}}

\newcommand{\prop}{\pi}
\newcommand{\Next}{\mathsf{X}}
\newcommand{\Until}{\mathsf{U}}
\newcommand{\Always}{\mathsf{G}}
\newcommand{\Event}{\mathsf{F}}

\renewcommand{\epsilon}{\varepsilon}

\newcommand{\eg}{{e.g.,~}}

\title{\LARGE \bf
Decomposition of Multi-Agent Planning under Distributed Motion and Task LTL Specifications 
}

\author{Jana Tumova and Dimos V. Dimarogonas
\thanks{This work was supported by EU STREP RECONFIG, European Union’s Horizon 2020 Research and Innovation Programme under the Grant Agreement No.644128 (AEROWORKS), Swedish Research Council (VR), and H2020 European Reasearch Council (ERC) Starting Grant BUCOPHSYS. The authors are with the ACCESS Linnaeus Center, School of Electrical
Engineering, KTH Royal Institute of Technology, Stockholm,
Sweden and with the KTH Centre for
Autonomous Systems. Email:
        {\tt\small tumova@kth.se, }%
        {\tt\small dimos@kth.se}.}%
}

\begin{document}

\maketitle
\thispagestyle{empty}
\pagestyle{empty}

\begin{abstract}

The aim of this work is to introduce an efficient procedure for discrete multi-agent planning under local complex temporal logic behavior specifications. While the first part of an agent's behavior specification constraints the agent's trace and is independent, the second part of the specification expresses the agent's tasks in terms of the services to be provided along the trace and may impose requests for the other agents' collaborations. To fight the extreme computational complexity of centralized multi-agent planning, we propose a two-phase automata-based solution, where we systematically decouple the planning procedure for the two types of specifications. At first, we only consider the former specifications in a fully decentralized way and we compactly represent each agents' admissible traces by abstracting away the states that are insignificant for the satisfaction of their latter specifications. Second, the synchronized planning procedure uses only the compact representations. The satisfaction of the overall specification is guaranteed by  construction for each agent. An illustrative example demonstrating the practical benefits of the solution is included.

\end{abstract}

\section{INTRODUCTION}

Automated synthesis of correct-by-design controllers complying with complex behavior specifications has recently found a wide support in formal verification methods and temporal logics. A number of the suggested solutions build on a hierarchical approach\cite{hadas09TL, marius-tac2008, nok-hscc2010, kavraki-ram}, where the dynamics of the system is abstracted into a finite, discrete transition system, a discrete plan that meets the specification is synthesized and next translated into a controller for the original system.
This work focuses on a multi-agent version of the above problem, and particularly on the second step of the hierarchical approach. 
We consider a {heterogeneous} team of agents (e.g., robots), that are assigned a behavior specification each, comprising of the requirements on the sequence of states they go through (e.g., the robots' trajectories) and the services they provide (e.g., loading an object). The agents' tasks are though not completely independent; in order to achieve their task a collaboration of the other agents might be required (e.g., helping to load a heavy object). 
Our aim in this paper is to efficiently synthesize a strategy for each agent, such that all requirements are met. As a specification language, we use Linear Temporal Logic (LTL), for its resemblance to natural language~\cite{hadas-icra2012}, and expressive power. The goal is viewed from the bottom-up perspective, where we introduce the notion of local specification satisfaction based on the individual agents' own viewpoints. 

Multi-agent planning from temporal logic specification has been explored in several recent works. 
Planning for a team of robots from a global LTL specification was addressed, e.g., in~\cite{loizou-cdc2005,marius-cdc2011}.
 The authors in~\cite{sertac-ijnc2010} considered LTL  for vehicle routing and in~\cite{lygeros-ecc2013} for search and rescue missions. A decentralized control of a robotic team from local LTL specification with communication constraints is proposed in~\cite{dimos-cdc12, meng-cdc14}. In contrast to our work, the agents therein do not impose any constraints on the other agents' behavior. 
In~\cite{yushan-tr2012,alphan-ijrr2013}, a top-down approach to LTL planning is investigated; the team is given a global specification and an effort is made to decompose the formula into independent local specifications that can be treated separately. 
In~\cite{meng-ijrr2015}, bottom-up planning from LTL specifications is considered, and a partially decentralized solution is proposed that decomposes the group into clusters of dependent agents.
To cope with the state space explosion, in~\cite{ acc14, automatica} we have proposed a receding horizon approach to multi-agent planning.

In this paper, we take into account an extension of the previous problem, where besides the high-level \emph{tasks} in terms of services to be provided, the agents' behaviors are limited by additional temporal logic constraints, allowing to express safety, surveillance, sequencing, or reachability properties of their traces, which we call \emph{motion} specifications for simplicity.
We propose to decouple the planning for the former and the latter parts of the specification. The main idea is to build local synchronized product automata for each of the agents' motion separately and reduce their size before creating the global product. 
The contribution of this paper can thus be summarized as the introduction of an efficient, bottom-up control strategy synthesis for multi-agent systems from local LTL motion and task specifications. The designed strategy allows the agents to execute their finite plans to a large extent independently, in an asynchronous manner. 

{The rest of the paper is structured as follows. In Sec.~\ref{sec:prelims}, we fix necessary notation. Sec.~\ref{sec:pf} formalizes the problem statement. In Sec.~\ref{sec:solution}, the details of the solutions are provided. An illustrative example is given in Sec.~\ref{sec:experiment}. 

\section{NOTATION AND PRELIMINARIES}
\label{sec:prelims}
Given a set $\Set$, let 
$2^\Set$,
and $\Set^\omega$
denote 
the set of all subsets of $\Set$, 
and the set of all infinite sequences of elements of $\Set$, respectively. 
The $i$-th element of the sequence $w$ is $w(i)$, and a \emph{subsequence} of  an infinite sequence $w = w(1)w(2)\ldots$ is a finite or infinite sequence $w(\i_1)w(\i_2)\ldots$, such that $1 \leq \i_j \leq \i_{j+1}$, for all $1 \leq j$. 
Boolean operator XOR (exclusive or) is denoted by~$\oplus$, positive integers and non-negative real numbers are denoted by $\Nat$ and $\Real_0^+$, respectively.

\label{sec:prelims:synthesis}

A \emph{labeled transition system (TS)} is a tuple $\TS$, where
$S$ is a finite set of states;
$s_{init} \in S$ is the initial state;
$A$ is a finite set of actions;
{$T \subseteq S \times A \rightarrow S$} is a partial deterministic transition function;
$\Pi$ is a set of atomic propositions; and
$L: S \rightarrow 2^\Pi$ is a labeling function.
A \emph{trace} of $\T$ is an infinite alternating sequence of states and actions $\tau = s_1\alpha_1s_2\alpha_2\ldots$, such that $s_1=s_\init$, and for all $i \geq 1$, 
$T(s_i,\alpha) = s_{i+1}$.


An \emph{LTL formula} $\varphi$ over {the set of atomic propositions $\Ser$} is defined
  inductively:
  (i) $\prop \in \Ser$ is a formula, and
  (ii) if $\varphi_1$ and $\varphi_2$ are formulas, then $\varphi_1 \lor
    \varphi_2$, $\lnot \varphi_1$, $\Next\, \varphi_1$, $\varphi_1\,\Until\,\varphi_2$, $\Event \, \varphi_1$, and $\Always \, \varphi_1$
    are each {a formula},
 where $\neg$ and $\vee$
 are standard Boolean connectives, and $\Next$ (next), $\Until$ (until), $\Event$ (eventually), and  $\Always$ (always) are temporal operators.
The semantics of LTL are defined over infinite words over~$2^\AP$ (see, e.g.,~\cite{principles}).
A trace $\tau = s_1\alpha_1s_2\alpha_2\ldots$ of $\T$ \emph{satisfies}  an LTL formula $\varphi$ over $\Pi$ ($\tau \models\varphi$) iff $L(s_1)L(s_2) \ldots$ satisfies $\varphi$ ($L(s_1)L(s_2)\ldots \models\varphi$).

A \emph{B\"uchi automaton (BA)} is a tuple $\A =  (Q,q_{init},\Sigma,\delta,F)$, where
$Q$ is a finite set of states; 
$q_{init}\in Q$ is the initial state; 
$\Sigma$ is an input alphabet; 
$\delta \subseteq Q \times \Sigma \times Q$ is a non-deterministic transition relation; 
$F \subseteq Q$ is an accepting condition.
A \emph{run} of the BA $\B$ from a state $q_1 \in Q$ \emph{over} a word {$w=\sigma_1\sigma_2\ldots \in \Sigma^\omega$}  is a sequence
$\rho=q_1q_2\ldots$, such that 
{$(q_i,\sigma_i,q_{i+1}) \in \delta$}, for all $i\geq 1$ and it is \emph{accepting} if it intersects $F$ infinitely many times. A word $w$ is \emph{accepted} by $\B$ if there exists an accepting run over~$w$ from the state $q_\init$. 
Any LTL formula $\varphi$ over $\Pi$ can be algorithmically translated into an equivalent BA $\B$ with $\Sigma=2^\Pi$. 
Any BA can be also viewed as a graph and the terminology from graph theory applies.



\section{PROBLEM FORMULATION}
\label{sec:pf}
Two general viewpoints can be taken in multi-agent planning: either the system acts as a team with a common goal, or the  agents have their individual, local specifications. Here, we adopt the second viewpoint and we do not look at the global team behavior. In contrast,  we
propose local specification satisfaction to determine whether an agent's task is fulfilled from its own perspective.
Each agent's temporal specification comprises of a \emph{motion}  
and a \emph{task} formula. 
Loosely speaking, the motion formula can be viewed as the agent's safety, reachability, surveillance or sequencing, while the task one represents its {effectiveness}, i.e, the motion formula limits the sequence of states the agent's goes through, while  the task formula corresponds to an accomplishment of a high-level task composed of simple services that the agent has the ability to provide. Unlike the motion one, the task formula may involve collaboration requirements.

Our goal is to synthesize a control strategy for each agent that involves (i)~the actions to be executed and (ii)~the requests for synchronization imposed on other agents. Together, the strategies have to ensure the local satisfaction of each of the motion and task specifications regardless of the time duration of the planned action executions.


\subsection{System Model}

{Let us consider $N$ agents (e.g., robots in a partitioned environment). Each agent $i \in \N$, where $\N = \{1,\ldots, N\}$ has two different types of capabilities: the ability to execute an action primitive, and to synchronize with the others. }


\subsubsection{Action Execution Capabilities} The agent $i$'s action-execution capabilities are modeled as a  finite TS $\TSi$. The states of the TS correspond to states of the agents (\eg the location of the robots in the regions of the environment). The atomic propositions represent inherent properties of the system states (\eg the robot is in a safe region). The actions represent abstractions of the agent's low-level controllers, and the transitions between the states correspond to the agent's capabilities to execute the actions (\eg the ability of the robots to move between two regions of the environment). The traces are, roughly speaking, the abstractions of the agents' long-term behaviors (\eg the robots' trajectories). 
Each of the agents' action executions takes a certain amount of time. Given a trace $\tau = s_1\alpha_1s_2\alpha_2\ldots$ of $\T_i$, we denote by $\Delta_{\alpha_j} \in \Real_0^+$ the time \emph{duration} of the transition $s_j\xrightarrow{\alpha_j}s_{j+1}$. Note that a transition duration is arbitrary and unknown prior its execution, and that the execution of the same action $\alpha$ may induce different transition durations in its different instances on the trace. 

\subsubsection{Synchronization Capabilities} Next to the action-execution capabilities, the agents have the ability to \emph{synchronize}, i.e., to wait for each other and to proceed with the further  execution simultaneously. The synchronization is modeled through the \emph{synchronization requests}. While being in a state $s$, an agent $i$ can send a request $sync_i(I)$ to a subset of agents $\{i\} \subseteq I \subseteq \N$ notifying that it is ready to synchronize. Then, before proceeding with the execution of any action $\alpha \in A_i$, it has to wait till $sync_{i'}(I)$ has been sent by each agent $i' \in I$, i.e., till the moment when each agent $i' \in I$ is ready to synchronize, too. For simplicity, we assume the perfect propagation of the synchronization requests. The synchronization is immediate once each of the agents $i' \in I$ has sent its request  $sync_{i'}(I)$ and all agents in $I$ start executing the next action collaboratively, at the same time. 
The set of all synchronization requests of an agent $i$ is 
$\Sync_i = \{\sync_i(I) \mid \{i\} \subseteq I \subseteq \N\}$.
Note that the synchronization request $sync_i(\{i\})$ indicates that no synchronization with the others is needed, and the next action of agent $i$ is executed immediately, independently on the other agents.
We assume that each agent sends a synchronization request instantly once it completes an action execution and that it starts executing an action instantly once it synchronizes with the other agents. 
Instead of allowing for idling, we include an existence of a special action $\stay_i \in A_i$ and a so-called \emph{self-loop} $s \xrightarrow{\stay_i} s$ for all $s \in S_i$, and all $\T_i$, where $i \in \N$.
Given a trace $\tau_i = s_{i,1}\alpha_{i,1}s_{i,2}\alpha_{i,2}\ldots$ of $\T_i$,  we denote by $\Delta_{s_{i,j}} \in \Real_0^+$ the time duration of the synchronization that has been requested in the state $s_{i,j}$. 
Note that $\Delta_{s_{i,j}} = 0$ for $\sync_i(\{i\})$ in $s_{i,j}$.

\subsubsection{Services} Each of the agents' specification is given via two components: the first one are temporal requirements on the atomic propositions that need to hold along its trace, and the second one is a task given in terms of events of interest associated with their actions, which we call \emph{services} (e.g., an object pick-up). The set of services that can be \emph{provided} by an agent $i\in \N$ is $\Ser_i$. Services are {provided} within the agents' transitions; each action $\alpha \in A_i$ is associated, or in other words \emph{labeled}, either with 
(i) a service set $\sigma \in 2^{\Ser_i}$ provided by $i$  upon the execution of $\alpha$, or 
(ii) a special \emph{silent} service set $\Epsilon_i = \{\epsilon_i\}$, where $\epsilon_i \not \in {\Ser_i}$ indicating that the action $\alpha$ does not provide any event of interest.
{Hence, two additional components of the agent $i$'s model are
the set of all available services $\Ser_i$ and the action-labeling function $\Lab_i: A_i \to 2^{\Pi_i} \cup 2^{\Epsilon_i}$.} Note that we specifically distinguish between a silent service set $\Epsilon_i$ and an empty service set $\{\}$. 
The above mentioned self-loops of form $s \xrightarrow{\stay_i} s$ that we use to model the agents' ability of staying in their  respective current states are labeled with the silent service set $\Lab_i(\stay_i) = \Epsilon_i$.
Without loss of generality, we assume that $\Ser_i \cap \Ser_{i'} = \emptyset$, for all $i,i' \in \N$, such that $i\neq i'$, which is necessary for the final step of our solution. 
Finally, we model an agent $i \in \N$ as  the tuple 
$\model_i = (\TSi, \Sync_i, \Ser_i, \Lab_i)$.

\subsubsection{Behaviors} \emph{Behavior} of an agent $i$ is defined by its states,  actions,  synchronizations with the others, and the time instants of the action executions and the synchronizations.

\begin{definition}[Behavior and strategy] A behavior of an agent $i$ is a tuple $\beta = (\tau,\gamma, \mathbb T)$, where
\begin{itemize}
\item 
$\tau = s_1\alpha_1s_2\alpha_2\ldots$ is a trace of $\T_i$;
\item 
$\gamma = r_1r_2\ldots $ is a \emph{synchronization sequence}, where $r_j \in {\Sync_i}$ is the synchronization request sent at $s_j$; and
\item 
$\mathbb{T} = t_{s_1}t_{\alpha_1}t_{s_2}t_{\alpha_2}\ldots$ is a non-decreasing \emph{behavior time sequence}, where  
$t_{s_j}$ is the time instant when the synchronization request $r_j$ was sent, and $t_{\alpha_j}$ is the time instant when the action $\alpha_j$ started being executed. The following properties hold: $t_{s_1}=0$, and for all $j \geq 1$, $t_{s_{j+1}}  - t_{\alpha_j} = \Delta_{\alpha_j}$, and $t_{\alpha_j} - t_{s_j}=\Delta_{s_j}$.
\end{itemize}
{A \emph{strategy} $(\tau,\gamma)$ for an agent $i$ is a trace $\tau$ and a synchronization sequence $\gamma$.} 
\label{def:behavior}
\end{definition}

{
The notion of behavior does not reflect the above described inter-agent synchronization rules. To that end, we define 
\emph{behaviors induced by strategies} admissible by a team of agents.} 
We denote the behavior of an agent $i \in \N$ by $\beta_i = (\tau_i,\gamma_i,\mathbb T_i)$, and we use $\tau_i = s_{i,1}\alpha_{i,1}s_{i,2}\alpha_{i,2}\ldots$, $\gamma_i = r_{i,1}r_{i,2}\ldots $, and $\mathbb{T}_i = t_{s_{i,1}}t_{\alpha_{i,1}}t_{s_{i,2}}t_{\alpha_{i,2}}\ldots$ to denote its trace, synchronization and time sequence. 

\begin{definition}[Induced behaviors]
The set of behaviors induced by a collection of strategies $(\tau_1, \gamma_1), \ldots, (\tau_N, \gamma_N)$ of  agents in $\N$ are the subset of the collections of their behaviors $\mathbb B \subseteq \{\mathfrak B \in \{ \beta_1,\ldots,\beta_N \mid \beta_i \text{ is a behavior of agent } i\}\}$ satisfying the following condition for all $i \in \N$, and $j \geq 1$: Suppose that $r_{i,j} = sync_i(I)$. Then for all $i' \in I$ there exists a \emph{matching index} $j' \geq 1$, such that $r_{i',j'} = sync_{i'}(I)$, and $t_{\alpha_{i,j}} = t_{\alpha_{i',j'}}$. Furthermore, there exists at least one $i' \in I$, such that $t_{s_{i',j'}} = t_{\alpha_{i',j'}}$, i.e., such that $\Delta_{s_{i',j'}} = 0$, for the matching index $j'$.
\label{def:compatible}
\end{definition}


\subsection{Motion and Task Specifications}

{The individual agents' tasks are collaborative; on the one hand, they involve requirements on the agents' states, i.e., the atomic propositions associated with them and on the other hand, they concern the respective agent's services as well as the services of the other agents. Formally,} each of the agents is given 
(i) an \LTLX formula $\phi_i$ over $\AP_i$, which we call a \emph{motion} specification and (ii) a collaborative LTL formula
$\psi_i$ over
{$\Sers = \bigcup_{i' \in \N} \Ser_{i'}$}, which we call a \emph{task} specification. 
{Consider for a moment a single agent $\model_i = (\T_i,\Sync_i, \Ser_i, \Lab_i)$, and its behavior $\beta = (\tau, \gamma, \mathbb T)$, where $\tau = s_{1}\alpha_{1}s_{2}\alpha_{2}\ldots$, and  $\mathbb{T} = t_{s_1}t_{\alpha_1}t_{s_2}t_{\alpha_2}\ldots$.

\begin{definition}[Words and time sequences] We denote by  $v_\tau = \varpi_1\varpi_2\ldots = \Lab_i(\alpha_{1})\Lab_i(\alpha_{2})\ldots  \in (2^{\Ser_i} \cup 2^{\Epsilon_i})^\omega$ the  \emph{service set sequence} associated with $\tau$. 
The \emph{word} $w_\tau$ produced by $\tau$ is the subsequence of the \emph{non-silent} elements of $v_\tau$; $w_\tau = \varpi_{\i_1}\varpi_{\i_2}\ldots \in (2^{\Sigma_i})^\omega$, such that $\varpi_1, \ldots, \varpi_{\i_1-1} = \Epsilon_i$, and for all $j \geq 1$, $\varpi_{\i_j} \neq \Epsilon_i$ and $\varpi_{\i_j+1}, \ldots, \varpi_{\i_{j+1}-1} = \Epsilon_i$.
With a slight abuse of notation, we use $\mathbb{T}(v_\tau) = t_1t_2\ldots = t_{\alpha_1}t_{\alpha_2}\ldots$ to denote the \emph{service time sequence}, i.e. the subsequence of $\mathbb T$ when the services are provided.
Furthermore, $\mathbb{T} (w_\tau)= t_{\i_1}t_{\i_2}\ldots$ denotes the \emph{word time sequence}, i.e. the subsequence of $\mathbb{T}(v_\tau)$ that corresponds to the time instances when the non-silent services are provided. 
\end{definition}
}

As in this work we are interested in infinite behaviors, we  consider as \emph{valid} traces only those producing infinite words.

\begin{definition}[Service set at a  time]
Let  $\tau$ be a trace of $\T_i$ with $v_\tau = \varpi_1\varpi_2\ldots$, and 
$\mathbb{T}(v_\tau) = t_{1}t_{2}\ldots$. Given $t \in \Real_0^+$, 
the service set $v_\tau(t) \in 2^{\Pi_i} \cup 2^{\Epsilon_i}$ provided at time $t$ is
$ v_\tau(t) = 
  \varpi_j \text{ if $t=t_j$ for some $j \geq 1$; and }
    v_\tau(t) =  \Epsilon_i \text{ otherwise.}$
 
\end{definition}

Note that the above definition is designed in a way accommodating the general asynchronicity of the agents.
Each formula $\psi_i$ is interpreted locally, from the agent $i$'s point of view, based on the word {$w_{\tau_i}$} it produces and on the services of the other agents $i' \in \N$ provided at the time instances $\mathbb T(w_{\tau_i})$. In other words, the agent $i$  takes into consideration the other agents' services only at times, when $i$ provides a non-silent service set (even an empty one) itself.

\begin{definition}[Local LTL satisfaction]
Let $\mathfrak B = \beta_1,\ldots,\beta_N$ be a collection of behaviors, where $\beta_{i} = (\tau_i,\gamma_i,\mathbb T_i)$ for all $i \in \N$ and let $\mathbb T(w_{\tau_i}) =  t_{i,\i_1}t_{i,\i_2}\ldots$ be the word time sequence of agent $i$. The \emph{local word} produced by $\mathfrak B$ is
$w_{\mathfrak{B_i}} = \omega_{i,\i_1}\omega_{i,\i_2}\ldots, \text{ where}$
$\omega_{i,\i_j} = \big(\bigcup_{i' \in \N} v_{\tau_{i'}}(t_{i,\i_j})\big) \cap  {\Sers}, \text{ for all } j\geq 1$.
$\mathfrak{B}$ \emph{locally satisfies} the formula $\psi_i$ for the agent~$i$, $\mathfrak{B} \models_i \psi_i$, iff $w_{\mathfrak{B}_i} \models \psi_i$.\label{def:localsat}
\end{definition}

Note that even if $\psi_i = \psi_j$,
it is possible that $\mathfrak{B} \models_i \psi_i$ for the agent $i$, while $\mathfrak{B} \not \models_j \psi_j$ for the agent $j$, since it might be the case that $w_{\mathfrak{B}_i} \neq w_{\mathfrak{B}_j}$.


\subsection{Problem Statement}
\begin{problem}
\label{prob:main}
\emph{Consider} a set of agents $\N = \{1,\ldots,N\}$, and suppose that each agent $i \in \N$ is modeled as a tuple $\model_i = (\TSi, \Sync_i, \Ser_i, \Lab_i)$, 
and assigned a task in the form of an \LTLX formula $\phi_i$ over $\Pi_i$ and $\psi_i$ over 
$\Sers =  \bigcup_{i' \in \N} \Ser_{i'}$.
\emph{For each $i \in \N$ find a strategy, i.e.,}
(i)~a trace $\tau_i = s_{i,1}\alpha_{i,1}s_{i,2}\alpha_{i,2}\ldots$ of $\T_i$ and 
(ii)~a synchronization sequence $\gamma_i$ over $\Sync_i$
with the property that the set of induced behaviors $\mathbb B$ from Def.~\ref{def:compatible} is nonempty, and for all $\mathfrak B \in \mathbb B$  and all $i \in \N$, it holds that 
the trace $\tau_i$ satisfies $\phi_i$
and the word $w_{\mathfrak B_i}$ produced by $\mathfrak B$ locally satisfies $\psi_i$ for the agent $i$ in terms of Def.~\ref{def:localsat}.
\end{problem}

As each LTL formula can be translated into a BA, from now on, we pose the problem  with the motion specification of each agent $i$  given as a  BA $\B_i^\phi = (Q_i^\phi, q_{\init, i}^\phi, \delta_i^\phi, 2^{\AP}, F_i^\phi)$, and the task one as a BA $\B_i^\psi = (Q_i^\psi, q_{\init, i}^\psi, \delta_i^\psi, 2^{\Sers}, F_i^\psi)$. Given that $\tau_i = s_{i,1}\alpha_{i,1}s_{i,2}\alpha_{i,2}\ldots$, the motion specification satisfaction is formulated as  $L(s_{i,1})L(s_{i,2})\ldots \in \mathit{Lang(\B_i^\phi)}$ and the local satisfaction of  the task specification as $w_{\mathfrak B_i} \in \mathit{Lang(\B_i^\psi)}$.



\section{PROBLEM SOLUTION}
\label{sec:solution}



Even though the agents' motion specifications are mutually  independent, each of them is dependent on the agent's respective task specification, which is dependent on the task specifications of the other agents. As a result, the procedures of synthesizing the desired $N$ strategies cannot be decentralized in an obvious way. However, one can quite easily obtain a centralized solution when viewing the problem as a synthesis of a single team strategy.

First of all, we realize that if there exists a solution to Problem~\ref{prob:main}, then there exists a solution in which all the agents synchronize after every single step. Informally, the reason is that each agent $i$ is able to execute action $\stay_i$, i.e., that $L_i(\stay_i) = \Epsilon_i$, and that the motion specification $\phi_i$ is a formula of \LTLX, i.e., that the execution of $\stay_i$ does not influence its satisfaction.
The standard automata-based procedure for solving the LTL control strategy synthesis problem for deterministic systems 
can  be then easily applied in multi-agent settings in a centralized way.
1) A synchronized TS $\T_\N$ with the set of states $S_\N = S_{1}\times \ldots \times S_{N}$ is built that represents all stepwise-synchronized traces of the agents in $\N$. 
2) A BA $\B_\N$ is built for the formula $\bigwedge_{i \in \N} (\phi_i \wedge \psi_i)$ over the alphabet ${2^{\bigcup_{i\in \N} \AP_i \cup \Sers}}$.
3) A product automaton $\P_\N$ of $\T_\N$ and $\B_\N$ is constructed that captures only the agents' traces that are admissible by $\T_\N$ and result into behaviors that satisfy $\phi_1,\ldots,\phi_N$, and locally satisfy $\psi_{1},\ldots,\psi_{N}$, for the agents $1,\ldots, N$, respectively. The product $\P_\N$ is analyzed using standard graph algorithms in order to find its accepting run that projects onto the desired traces of $\T_i$, for all $i \in \N$.
4) The synthesized strategy for an agent $i \in \N$ is a trace $\tau_i = s_{i,1}\alpha_{i,1}s_{i,2}\alpha_{i,2}\ldots$ of $\T_i$, and a synchronization sequence $\gamma_i = r_{i,1}r_{i,2}\ldots$ where $r_{i,j} = \sync_i\{\N\}$, for all $j \geq 1$. We notice that the amount of synchronization required by agent $i \in \N$ can be reduced by replacing $r_{i,j} = \sync_i(\N)$ with $\sync_i(\{i\})$, if for each agent $i' \in \N$ it holds that $\Lab_{i'}(\alpha_{i',j}) = \Epsilon_{i'}$, since in such a case, $w_{\mathfrak B_i}$ remains unchanged. Furthermore, the $\stay_i$ actions can be removed from $\tau_i$ without any harm, since they influence neither the satisfaction of $\phi_i$, nor the word $w_{\mathfrak B_i'}$ for any $i'\in \N$.

\subsection{Our solution}
A major drawback of the centralized solution is the state space explosion, which makes it practically unusable. We aim to decentralize the solution as much as possible. Namely, we aim to separate the synthesis of service plans yielding the local satisfaction of the task specifications from the syntheses of traces that guarantee the motion specifications. 
Our approach is to pre-compute possible traces and represent them efficiently, abstracting away the features that are not significant for the synthesis of action plans. This abstraction serves as a guidance for the action and synchronization planning, which, by construction, allows for finding a trace complying with both the synthesized action and synchronization plans and and the motion specification. 


\subsubsection{Preprocessing the motion specifications}
Consider for now an agent $i \in \N$ modeled as $\model_i = (\T_i, \Sync_i, \Ser_i, \Lab_i)$, where $\TSi$ and its motion specification BA $\B_i^\phi$. We slightly modify the classical construction of a product automaton of $\T_i$ and $\B_i^\phi$ to obtain a BA that represents the traces of $\T_i$ accepted by $\B_i^\phi$ and furthermore explicitly captures the services provided along the trace.

\begin{definition}[Motion product]
\label{def:PA}
The \emph{motion product} of a TS $\TSi$, and a BA $\B_i^\phi = (Q_i^\phi, q_{\init, i}^\phi, \delta_i^\phi, 2^{\AP_i}, F_i^\phi)$ is a BA $\P_i = (Q_{i}, q_{\init,i}, \delta_{i}, 2^{\Ser_i} \cup \{\Epsilon_i\}, F_{i})$, where
$Q_{i} = S_i \times Q_i^\phi$;
$q_{\init,i} = (s_{\init,i},q_{\init,i}^\phi)$;
$((s,q),\Lab_i(\alpha),(s',q') \in \delta_{i}$
if and only if $(s,\alpha,s') \in T_i$, and $(q,L_i(s),q') \in \delta_i^\phi$; and
$F_{i} = \{(s,q) \mid q \in F_i^\phi\}$.
\label{def:motion}
\end{definition}

Intuitively, we propose to keep only the significant states that have an outgoing transition labeled with $\Lab_i(\alpha) \neq \Epsilon_i$ and replace exact sequences of transitions in $\P_i$ with single transitions representing reachability. While doing so, we have to take into account whether the removed state is accepting or not to correctly preserve the accepting condition of $\P_i$.

\begin{definition}[Insignificant states in $\P_i$]
A state $p$ of the motion product $\P_i$ is \emph{significant} if it is (i) the initial state $p = q_{\init,i}$, or (ii) if there exists a transition $(p,\sigma,p') \in \delta_i$, such that $\sigma \neq \Epsilon_i$; and \emph{insignificant} otherwise.
\end{definition}

A \emph{reduced motion product} $\ddot \P_i$ is built  from $\P_i$ according to Alg.~\ref{alg:removal}. First, we remove all  insignificant non-accepting states and their incoming and outgoing transitions (lines~\ref{line:5},~\ref{line:6}). We replace each state with a set of transitions leading directly from the state's predecessors to its successors, i.e., we concatenate the incoming and the outgoing transitions (line~\ref{line:7}). The labels of the new transitions differ: if both labels of the concatenated incoming and outgoing transition are $\Epsilon_i$, then the new label will be  $\irrelevant$ to indicate that the transition represents a sequence of actions that are not interesting with respect to the local satisfaction of task specifications. On the other hand, if the label $\sigma$ of the incoming transition belongs to $2^{\Ser_i}$, we use the action $\sigma$ as the label for the new transition. Each path between two significant states in $\P_i$ maps onto a path between the same states in $\ddot \P_i$ and the sequences of non-silent services read on the labels of the transitions of the two paths are equal;
 and vice versa. 
Second, we handle the insignificant accepting states (line~\ref{line:8}) similarly to the non-accepting ones, however, we do not remove the states whose predecessors include a significant state in order to preserve the accepting condition. There is a correspondence between the infinite runs of $\P_i$ and the infinite runs of $\ddot \P_i$: for each run of $\P_i$ there exists a a run of $\ddot \P_i$, such that the states of the latter one are a subsequence of the states of the former one, the sequences of non-silent services read on the labels of the transitions of the two runs are equal, and that the latter one is accepting if and only if the former one is accepting; and vice versa. This correspondence will allow us to reconstruct a desired run of $\P_i$ from a  run of $\ddot \P_i$, as we will show in Sec.~\ref{sec:synthesis}.
From $\ddot \P_i$, we further remove all states from which none of the  states in $\ddot F_i$ is  reachable, and we can keep only one copy of duplicate states that have analogous incomming and outgoing edges (omitted from Alg.~\ref{alg:removal}).


\vspace{-0.3cm}
\begin{algorithm}
\caption{Reduction of the motion product}
\begin{algorithmic}[1]
\small{
\INPUT motion product $\P_i = (Q_{i}, q_{\init,i}, \delta_{i},  2^{\Ser_i} \cup 2^{\Epsilon_i}, F_{i})$
\OUTPUT reduced BA  $\ddot{\P}_i = (\ddot Q_{i}, \ddot q_{\init,i}, \ddot \delta_{i},2^{\Ser_i} \cup 2^{\Epsilon_i}, \ddot F_{i})$
\STATE initialize $\ddot \P_i  := \P_i$
\FORALL {insignificant states $p \in Q_i \setminus F_i$}
\STATE $\ddot Q_i  := \ddot Q_i \setminus \{p\}$ ~\label{line:5}
\STATE $\ddot \delta_i := \ddot \delta_i \setminus 
\{(p',\sigma,p),(p,\sigma,p') |  p' \in \ddot Q_i, \sigma \in 2^{\Sigma_i} \cup 2^{\Epsilon_i}\} $\label{line:6}
\STATE $\ddot \delta_i := \ddot \delta_i \cup 
\{ (p',\sigma, p'') \mid p',p'' \in \ddot Q_i, \sigma \in 2^{\Sigma_i} \cup 2^{\Epsilon_i}, $ $ (p',\sigma,p), (p,\Epsilon_i,p'') $ $ \in \ddot \delta_i\}$\label{line:7}
\ENDFOR
\FORALL {insignificant states $p \in F_i$, such that all predecessors of $p$ in $\ddot \P_i$ are insignificant}  \label{line:8}
\IF {$(p,\Epsilon_i,p) \in \ddot \delta_i$}
\STATE $\ddot \delta_i := \ddot \delta_i \cup \{(p',\irrelevant,p') \mid  (p',\Epsilon_i,p) \in \ddot \delta_i \}$
\ENDIF
\STATE $\ddot Q_i  := \ddot Q_i \setminus \{p\}$ ~\label{line:12}
\STATE  $\ddot \delta_i := \ddot \delta_i \setminus 
\{(p',\sigma,p),(p,\sigma,p') |  p' \in \ddot Q_i, \sigma \in 2^{\Sigma_i} \cup 2^{\Epsilon_i}\} $\label{line:13}
\STATE $\ddot \delta_i := \ddot \delta_i \cup  \{ (p',\Epsilon_i, p'') \mid p',p'' \in \ddot Q_i, (p',\Epsilon_i,p), $ $ (p,\Epsilon_i,p'') $ $ \in \ddot \delta_i\}$
 \label{line:14}
\ENDFOR
}
\end{algorithmic}
\label{alg:removal}
\end{algorithm}
\vspace{-0.3cm}




\subsubsection{Preprocessing the task specifications}
Next, we build a local task and motion product automaton for each agent $i$ separately, to capture the admissible traces of $i$ that comply both with its motion and task specification. At this stage, the other agents' collaboration capabilities are not yet included.

\begin{definition}[Task and motion product]
\label{def:PA}
The \emph{task and motion product} of a reduced motion product automaton $\ddot \P_i = (\ddot Q_{i}, \ddot q_{\init,i}, \ddot \delta_{i}, 2^{\Sigma_i} \cup 2^{\Epsilon_i}, \ddot F_{i})$, and the task specification BA $\B_i^\psi = (Q_i^\psi, q_{\init, i}^\psi, \delta_i^\psi, 2^{\Sers}, F_i^\psi)$ is a BA $\bar \P_i = (\bar Q_{i}, \bar q_{\init,i}, \bar \delta_{i}, 2^{\Sers} \cup 2^{\Epsilon_i}, \bar F_{i})$, where
$\bar Q_{i} = \ddot Q_i \times Q_i^\psi \times \{1,2,3\}$;
$\bar q_{\init,i} = (q_{\init,i},q_{\init,i}^\psi,1)$; $\bar F_{i} = \{(q_1,q_2,2) \mid q_2 \in F_i^\psi\}$; and
$((q_1, q_2, j),\sigma,(q_1', q_2', j') \in \bar \delta_{i}$
iff
\begin{itemize}
\item[$\circ$] $\sigma = \Epsilon_i$, $(q_1,\Epsilon_i,q_1') \in \ddot \delta_i$, and $q_2 = q_2'$; or
\item[$\circ$] $\sigma \in 2^\Sers$,  $(q_1,\sigma \cap 2^{\Ser_i},q_1') \in \ddot \delta_i$, and $(q_2,\sigma,q_2') \in \delta_i^\psi$,
\end{itemize}
and
$j'=2$, if $j=1$ and $q_1' \in \ddot F_i$, 
$j'=3$ if $j=2$ and $q_2' \in \ddot F_i^\psi$,
$j'=1$ if $j=3$, and 
$j = j'$ otherwise.
\label{def:action}
\end{definition}

The above construction leverages ideas from building a BA that accepts a language intersection of multiple given BAs. The accepting runs of $\bar \P_i$ map onto accepting runs of $\ddot \P_i$, and hence to the traces of $\T_i$ satisfying the motion specification $\phi_i$, and onto accepting runs of $\B_i^\psi$; and vice versa. 
Clearly, some of the transitions of $\bar \P_i$ require a collaboration of some other agents, while some of them do not. In order to further reduce the size of $\bar \P_i$, we introduce a mapping $\Dep_i: \bar \delta_i  \rightarrow 2^\N$ to indicate for each transition $\trans= (p,\sigma,p') \in \bar \delta_i$ the set of agents $\Dep_i(\trans) \subseteq \N$ whose collaboration is required. We formalize $\Dep_i$ through the notion of assisting services. Moreover, for each agent $i$, we find a subset of its services that affect the local satisfaction of a task specification of another agent through the notion of globally {assisting} services.

\begin{definition}[Assisting services]
Suppose that $i \neq i'$. A service $\rho \in {\Ser_{i'}}$ is \emph{not assisting on a transition $(p,\sigma,p') \in  \bar \delta_{i}$} of $\bar \P_{i}$ if and only if it holds  that $(p,\sigma \cup \{\rho\} ,p') \in \bar \delta_{i} \iff (p,\sigma \setminus \{\rho\},p') \in \bar \delta_{i}$; it is \emph{assisting on $(p,\sigma,p')$} otherwise. 
$\Dep_i (\trans) =  \{i\} \, \cup \, \{i' \mid i'\neq i \text{ and } \exists \rho \in \Sigma_{i'} \text{ assisting on } \trans \}.$
\end{definition}


\begin{definition}[Globally assisting services]
A service $\rho \in {\Ser_{i'}}$ of agent $i' \in \N$ is \emph{globally assisting} if there exists $i \in \N$, $i \neq i'$, and a transition $\trans \in \bar \delta_{i}$, such that $\rho$ is assisting on $\trans$. 
\end{definition}


Now, similar ideas as in Alg.~\ref{alg:removal} can be used to reduce the size of each action and motion product $\bar P_i$, with the following, altered definition of significant states:

\begin{definition}[Insignificant states in $\bar \P_i$]
A state $p$ of the task and motion product $\bar \P_i$ is \emph{significant} if it is either (i)~the initial state $p = \bar q_{\init,i}$, or (ii) if there exists a transition $(p,\sigma,p') \in \bar \delta_i$, such that there exists a globally assisting service $\rho \in \sigma \, \cap \, {\Sigma_{i}}$, or (iii) if there exists a transition $(p,\sigma,p') \in \bar \delta_i$,  such that $\Dep_i(p,\sigma,p') \neq \{i\}$;
the state $p$ is \emph{insignificant} otherwise.
\end{definition}

{Informally, the condition (ii) states that at a significant state, the agent $i$ has the ability to assist other agents, whereas the condition (iii) states that at a significant state, the assistance of the other agents influences the agent $i$'s own task specification local satisfaction. 
{With a slight abuse of notation, we replace all labels of all outgoing transitions of insignificant states of $\bar P_i$ with $\Epsilon_i$. This time $\Epsilon_i$ represents a service set that is completely independent.
We can  now directly apply Alg.~\ref{alg:removal} to remove the insignificant states from $\bar P_i$, with the exception that $2^{\Sigma_i}$ is at all places replaced with $2^ \Sers$.
The resulting automaton is $\widehat \P_i = (\widehat Q_{i}, \widehat q_{\init,i}, \widehat \delta_{i}, 2^{\Sers} \cup 2^{\Epsilon_i}, \widehat F_{i})$.
Similarly as before, there is a correspondence beween the infinite runs of $\bar \P_i$ and the infinite runs of $\widehat \P_i$: for each run of $\bar \P_i$ there exists a a run of $\widehat \P_i$, such that the states of the latter one are a subsequence of the states of the former one, the sequences of services read on the labels of the transitions leading from the significant states of the two runs are equal, and that the latter one is accepting if and only if the former one is accepting; and vice versa. The number of states of $\widehat \P_i$ is at most twice the number of significant states in $\bar P_i$.
}





\subsubsection{Global product} 

From the reduced task and motion product automata $\widehat \P_1, \ldots, \widehat \P_N$, we build a single one, that finally represents the inter-agent collaborations. The construction is a twist to the well-known construction of BA intersection. We associate the transitions of the BA with the subsets of agents that are required to make this transition collaboratively, i.e., with the subsets of agents that need to synchronize prior this transition.

\begin{definition}[Global product]
\label{def:PA}
The \emph{global product} of the reduced task and motion product automata $\widehat \P_1, \ldots \widehat \P_N$, where $\widehat \P_i = (\widehat Q_{i}, \widehat q_{\init,i}, \widehat \delta_{i}, 2^{\Sers} \cup 2^{\Epsilon_i}, \widehat F_{i})$, for all $i \in \N$, is a BA $\P = (Q, q_{\init}, \delta, 2^{\Sers} \cup 2^{\Epsilon = \{\Epsilon_i \mid i \in \N\}}, F)$ with a mapping $\Dep: \delta \rightarrow 2^\N$, where $Q = \widehat Q_1 \times \ldots \times \widehat Q_N \times \{1,\ldots, N+1\}$; $ q_{\init} = (\widehat q_{\init,1}, \ldots, \widehat q_{\init,N},1)$; $\bar F_{i} = \{(q_1, \ldots ,q_N, N) \mid q_N \in \widehat F_N\}$; and
$\trans = ((q_1,\ldots,q_N,j),\sigma,(q_1', \ldots, q_N', j')) \in \delta$
iff either
\begin{itemize}
\item[$\circ$] $\exists i \in \N$, such that $\sigma = \Epsilon_i$, $(q_i, \Epsilon_i, q_i') \in \widehat \delta_i$, $q_{i'} = q_{i'}'$, for all $i' \neq i$, and 
$j' = j+1$ if $j=i$ and $q_i' \in\widehat F_i$, 
$j' = 1$ if $ j= N+1$, and
$j'  = j$ otherwise.
Then we set $\Dep(\trans) = \{i\}$; or
\item[$\circ$] $\sigma \in 2^\Sers$, and $\exists \, \I \subseteq \N$, such that for all $i\in \I$ it holds that  $(q_i,\sigma,q_i') \in \widehat \delta_i$ while for all $i \not \in \I$ it holds that $q_i = q_i'$. Moreover, $\bigcup_{i \in \I} \Dep_i(q_i,\sigma,q_i') \subseteq \I$, and
$j' = j+1$ if $j \in \I$ and $q_j' \in \widehat F_j$, 
$j' = 1$ if $ j= N+1$, and
$j' =j$ otherwise.
 Then we set $\Dep(\trans) = \I$.
\end{itemize}

\label{def:global}
\end{definition}

Each accepting run of the global product $\P$ maps directly on the accepting runs of the reduced task and motion product automata and vice versa, for each collection of accepting runs of the reduced task and motion product automata, there exists an accepting run of the global product $\P$. Note that by this construction, deadlocks are avoided.


\subsubsection{Synthesis of the strategies}
\label{sec:synthesis}
The final step of our solution is the generation of the strategy in $\P$ and its mapping onto a trace $\tau_i$ of $\T_i$ and a synchronization sequence $\gamma_i$ over $\Sync_i$, for all $i \in \N$.  Using standard graph algorithms (see, \eg~\cite{principles}), we find an accepting run $q_1 q_2 \ldots$ over a word $\sigma_1\sigma_2\ldots$ in $\P$, where $q_j = (\widehat q_{1,j},\ldots,\widehat q_{N,j}, k)$, for all $j \geq 1$. For each agent $i\in \N$:
\begin{itemize}
\item[(i)] Consider the sequence $\widehat q_{i,1} \widehat q_{i,2} \ldots$, that is obtained by the projection of the accepting run $q_1q_2\ldots$ onto the states of $\widehat \P_i$. Let $\iota_1 \iota_2\ldots$ be the subsequence of all indexes, such that $i \in \Dep (q_{\i_j},\sigma_{\i_j},q_{\i_j+1})$. Then $\widehat q_{i,\iota_1}\widehat q_{i,\iota_2}\ldots$ is an accepting run of $\widehat \P_i$ over the word $\sigma_{\i_1}\sigma_{\i_2}\ldots$ and $\widehat \gamma_i =  \widehat r_{i,1} \widehat r_{i,2} \ldots = \Dep (q_{\i_1}\sigma_{\i_1}q_{\i_1+1}) \Dep (q_{\i_2}\sigma_{\i_2}q_{\i_2+1})\ldots$ is a synchronization sequence.
\item[(ii)] From the construction of $\widehat P_i$, we know that there exists an accepting run $\bar q_{i,1} \bar q_{i,2}\ldots$ in $\bar \P_i$ over  a word $\bar \sigma_{i,1}\bar \sigma_{i,2}\ldots$, such that  $\widehat q_{i,\iota_1}\widehat q_{i,\iota_2}\ldots$ is a subsequence  $\bar q_{i,\ell_1}\bar q_{i,\ell_2}\ldots$ of  $\bar q_{i,1} \bar q_{i,2}\ldots$ and $\sigma_{\i_1}\sigma_{\i_2}\ldots$ is the corresponding subsequence $\bar \sigma_{i,\ell_1}\bar \sigma_{i,\ell_2}\ldots$ of  $\bar \sigma_{i,1}\bar \sigma_{i,2}\ldots$. The synchronization sequence $\bar \gamma_i = \bar r_{i,1} \bar r_{i,2} \ldots$ is constructed by setting $\bar r_{i,\ell_j} = \widehat r_{i,\i_j}$, for all $j \geq 1$, and $\bar r_{i,j} = \sync(\{i\})$ otherwise.
\item[(iii)] The mapping from the task and motion product $\bar P_i$ onto $\ddot \P_i$ is processed as follows: Suppose that $\bar q_{i,j} = (\ddot q_{i,j}, q_{i,j}^\psi,k)$, for all $j \geq 1$. The accepting run $\bar q_{i,\ell_1}\bar q_{i,\ell_2}\ldots$ then maps onto the accepting run $\ddot q_{i,\ell_1}\ddot q_{i,\ell_2}\ldots$ over $\ddot \sigma_{i,1}\ddot \sigma_{i,2}\ldots$, where $\ddot \sigma_{i,j} = \Epsilon_i$ if $\bar \sigma_{i,j} = \Epsilon_i$, and $\ddot \sigma_{i,j} = \bar \sigma_{i,j} \cap 2^{\Sigma_i}$ otherwise.
Naturally, $\ddot \gamma_i = \bar \gamma_i$.
\item[(iv)] The accepting run $q_{i,1}q_{i,1}\ldots$ of the motion product $\P_i$ over a word $\sigma_{i,1}\sigma_{i,1}\ldots$ and a synchronization sequence $\gamma_i$ is obtained from the reduced motion product $\ddot P_i$ by an analogous construction as in the case of mapping from the reduced task and motion product $\widehat \P_i$ onto the task and motion product $\bar \P_i$.
\item[(v)] Finally, an accepting  run $q_{i,1}q_{i,2}\ldots$ of $\P_i$ maps onto a trace $\tau_i$ of $\T_i$. The desired synchronization sequence is $\gamma_i$.
\end{itemize}

\subsection{Discussion}
\emph {Strategy execution and event-based recomputation:} In the above described solution, the strategy for each agent is constructed in an offline manner, and executed independently. However, the solution can be modified to an event-triggered one, where synchronization is triggered  by need and strategies are periodically recomputed in order to adapt to different execution speed of different agents and to handle various optimization criteria. The details of this solution is one of our future topics of interest.
\emph{Mutually dependency:} We supposed that all of the considered agents are mutually dependent through their task specifications, either directly or transitively. If this is not the case, the team of agents can be partitioned into dependency classes as suggested in~\cite{meng-ijrr2015}.
\emph{Complexity:} In the worst case, our solution meets the complexity of the centralized solution. However, this is often not the case. Since the size of the global product is highly dependent on the number of globally assisting services available in the agents' workspace, our solution is particularly suitable for systems with complex motion capabilities, sparsely distributed services of interest, and occasional needs for collaboration. Its benefits are demonstrated in Sec.~\ref{sec:experiment}.

\section{ILLUSTRATIVE EXAMPLE}
\label{sec:experiment}

An illustrative example with 3 robots is given in Fig.~\ref{fig:workspace}. Their workspace is partitioned into 100 cells, and a robot's state is defined by the cell it is present at. Agent 1 (green) is a ground vehicle and has to avoid the obstacles (in gray), while agents 2 (orange) and 3 (purple) are UAVs and their workspace is obstacle-free except for walls (thick).
The cells are each labeled with an atomic proposition from $\{\mathsf{R1, R2, R3, R4}\}$ indicating the room to which the cell belongs. Services are available in cell marked with stripes. Agent 1 can \emph{load} (green horizontal stripe) and \emph{unload} (green vertical). Agent 2 can \emph{help} (orange horizontal) or \emph{inform} (orange vertical). Agent 3 can \emph{assist} (purple horizontal).
The motion formulas are: For agent 1 to  avoid room $\mathsf{R1}$, $\phi_1 = \Always \neg \mathsf{R1}$, for agent 2 to  avoid $\mathsf{R2}$, $\phi_2 = \Always \neg \mathsf{R2}$, and for agent 3 to survey $\mathsf{R1}$ and $\mathsf{R2}$, $\phi_3 = \Always \Event \, \mathsf{R1} \wedge \Always \Event \, \mathsf{R2}$.
The task formulas are: For agent 1 to periodically load collaboratively with agents 2 and  3 and then to unload, collaboratively with agent 2 or  3: $\psi_1 = \mathit{load} \, \wedge \, \mathit{help} \,  \wedge \mathit{assist} \wedge \Always \,(\mathit{load} \Rightarrow \Next \, (\mathit{unload} \, \wedge \,(\mathit{help} \,\vee\, \mathit{assist}))) \wedge \Always\, (\mathit{unload} \Rightarrow \Next \, (\mathit{load} \, \wedge \, \mathit{help} \, \wedge \, \mathit{assist}))$, for agent 2 to periodically service $\mathit{inform}$, $\psi_2 = \Always \Event \, \mathit{inform}$, for agent 3 $\psi_3 = \mathit{assist} \, \vee \neg \mathit{assist}$, i.e., agent 3 is not assigned a specific task.
An example of as collection of the desired trajectories and synchronizations is given in Fig.~\ref{fig:workspace}.

\begin{figure}[!h]
\centering
\includegraphics[width=0.7\linewidth]{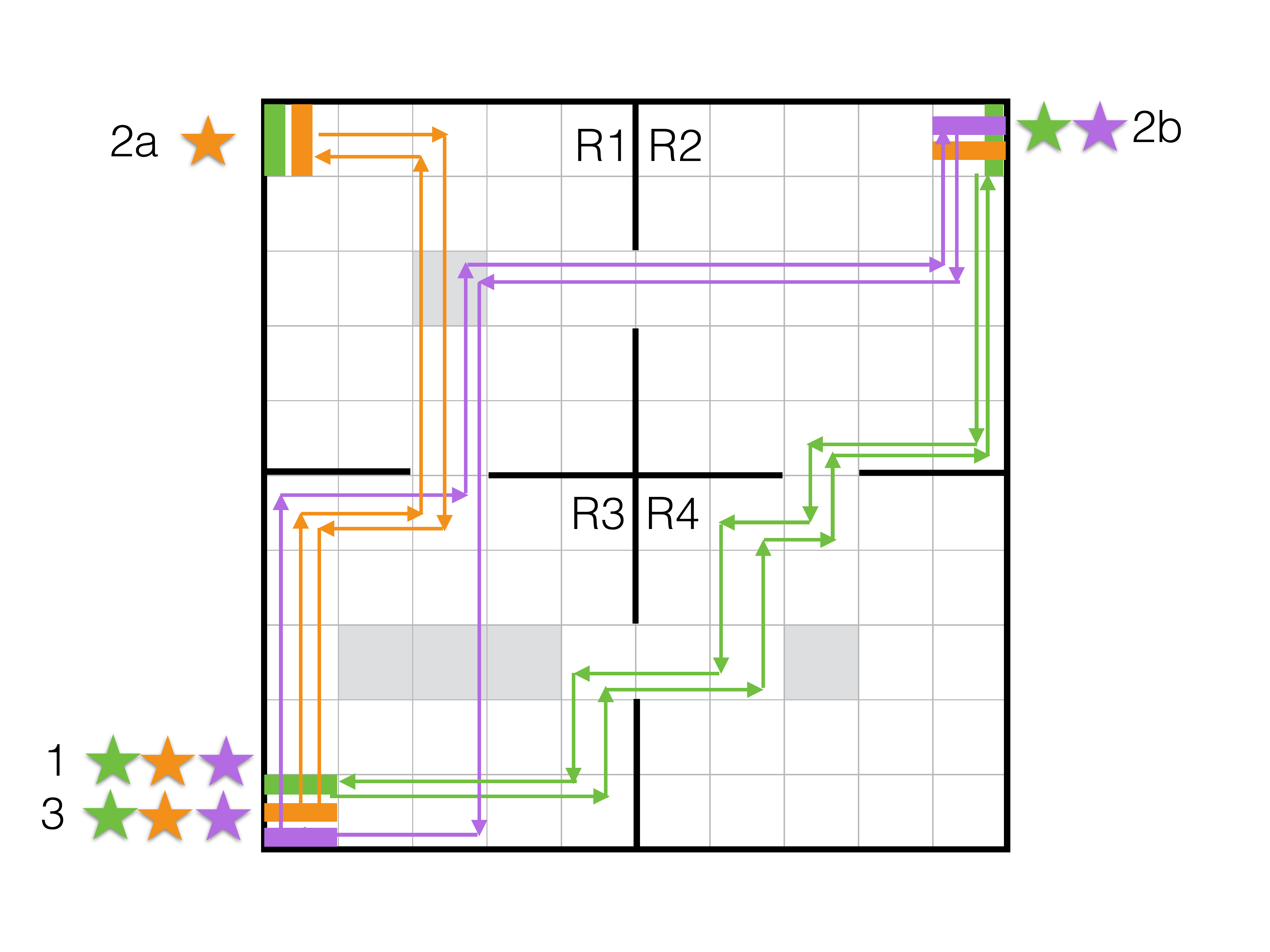}

\caption{\footnotesize{Examples of desired trajectories of 3 robots in a shared workspace. The synchronization events at the moments of providing nonsilent services are depicted with stars. First, all  agents synchronize in the bottom left corner and provide \emph{load, help, assist}; then agent 2 provides  \emph{inform} and then agents 1 and 3 provide  \emph{unload, assist}, or vice versa; finally, all  agents synchronize in the bottom left corner, and start periodically executing the above.}}

\label{fig:workspace}

\end{figure}

In classical centralized planning, only the synchronous product of the TSs of the three agents would have  $\approx$ 100 000 states. In our case, the BAs for the given formulas had 2,2,3,2,2,1 states, respectively, and hence their language intersection BAs reaches $2*2*3*2*2*1*7 \approx 330$ states. The overall product automaton has then up to $\sim 30$ millions states.
In contrast, in our solution, the individual task and motion products $\widehat \P_1, \widehat \P_2, \widehat \P_3$ after all reductions have 27,17, and 8 states, respectively. Hence the largest structure dealt with during the overall procedure, i.e., the product automaton $\P$ has only $27*17*8*4 \sim 15 000 $ states.




\bibliographystyle{plain}
\bibliography{refer}

\end{document}